\documentclass[pre,aps,showpacs,twocolumn,floatfix,superscriptaddress]{revtex4}

\usepackage{graphicx}
\usepackage{bm} 
\usepackage{epsfig}
\usepackage{color}
\usepackage{enumitem}
\usepackage[hidelinks]{hyperref}
\usepackage{amsmath}
\usepackage{float}%
\usepackage{hyperref}
\usepackage{siunitx}
\usepackage{booktabs}

\setlist[itemize]{leftmargin=*}
\setlist[enumerate]{leftmargin=*}	
\newcommand{\be}{\begin{equation}} 
\newcommand{\ee}{\end{equation}}
\newcommand{\bea}{\begin{eqnarray}}
\newcommand{\eea}{\end{eqnarray}}
\newcommand{\ba}{\begin{array}}
\newcommand{\ea}{\end{array}}

\begin{document}

\title{Effects of synapse location, delay and background stochastic activity on synchronising hippocampal CA1 neurons}

\author{Alessandro Fiasconaro}
\email{afiascon@unizar.es}
\affiliation{Dpto. de F\'isica de la Materia Condensada,
Universidad de Zaragoza. 50009 Zaragoza, Spain}
\affiliation{Instituto de Biocomputaci\'on y F\'isica de Sistemas
Complejos, Universidad de Zaragoza. 50018 Zaragoza, Spain}
\affiliation{Istituto di Biofisica, Consiglio Nazionale delle Ricerche. Palermo, Italy}

\author{Michele Migliore}
\email{migliore@cnr.it}
\affiliation{Istituto di Biofisica, Consiglio Nazionale delle Ricerche. Palermo, Italy}

\date{\today}

\begin{abstract}
We study the synchronisation of neurons in a realistic model under the Hodgkin-Huxley dynamics. To focus on the role of the different locations of the excitatory synapses, we use two identical neurons where the set of input signals is grouped at two different distances from the soma. The system is intended to represent a CA1 hippocampal neuron in which the synapses arriving from the CA3 neurons of the trisynaptic pathway appear to be localised in the apical dendritic region and are, in principle, either proximal or distal to the soma. Synchronisation is studied using a specifically defined spiking correlation function as a function of various parameters such as the distance from the soma of one of the synaptic groups, the inhibition weight and the associated activation delay. 
We found that the neurons' spiking activity depends nonmonotonically on the relative dendritic location of the synapses and their inhibitory weight, whereas the synchronisation measure always decreases with inhibition, and strongly depends on its activation time delay. The background activity on the somas results essentially independent on the fluctuation intensity and strongly support the importance of the balance between inhibition and excitation for neuronal synchronization.
\end{abstract}


\maketitle

\section*{Introduction.}
The mammalian hippocampus, located in the allocortex, is responsible for both short-term and long-term memory. It is the primary area affected in Alzheimer's disease and therefore plays an important role in cognitive function.

The CA1 pyramidal neurons receive two different sets of inputs: one coming directly from the entorhinal cortex (the so-called {\it direct path}) and the other from the CA3 neurons, which receive the signal from the granule cells, which in turn receive the signals from the entorhinal cortex (the {\it trisynaptic path}, so called because three sets of synapses are needed to complete this path).

Much of the brain's function is related to the synchronisation of incoming signals that ultimately lead to an appropriate output~\cite{2015Fries,1993Okeefe,1995Mainen,1995Hopfield}.
The modelling of of overlapping contributions in the depolarizing membrane potential is therefore of great importance for understanding the functioning of the brain and for developing theoretical approaches to its malfunctioning.

A large number of papers have investigated the possibility of synchronisation for a large scale of networks with different models. From Kuramoto oscillators~\cite{1975Kuramoto,2005Ritort}, the Izhikevich model~\cite{2003Izhikewich}, the integrated-and-fire (LIF) schemes~\cite{1999Abbott-LIF}, with its various generalisations~\cite{2005Brette,2021Gorski,2023Marasco}, or by using the Hodgkin-Huxley (HH) complete set of equations~\cite{1952HH}, as is the case here.

An overview on synchronization in the context of basic HH oscillators is given in the works by Boccaletti et al.\cite{2018Boccaletti, 2002Boccaletti} where two-neuron systems, as well as larger network oscillators, are presented.

This modelling effort has been elucidating many interesting synchronisation scenarios and developed a large comprehension of the synchronisation phenomenon. However, many of these approaches appear limited in different aspects: some of these models represent the neurons as single dynamic dots without internal structure. Therefore, no loss of energy due to the internal resistance is taken into account in this approach, nor delays due to the propagation of the current inside the neurons are included. Moreover, this kind of models present no relationship between their kinetic parameters and the biological magnitudes involved in neuron functioning.
Thus, besides the important basic knowledge it can offer on synchronisation phenomena, this mesoscopic approach neither creates a bridge between physical dynamics and biochemical understanding of the process, nor it can enter into possible medical treatments due to the lack of clear microscopic connections between the model parameters and the biological magnitudes (such as synaptic conductances, rise/decay times, signal propagation) that are modified by therapeutic treatments.

As regards the former case ---point-like models--- it is evident that the different ---and stochastic--- distribution of excitatory synapses in real neurons makes the spiking process a complex mechanism that involves the electrodynamics of pulse propagation along the dendrites, complicated by their shape and geometry~\cite{2015LiGirault}, or the channel types and their distributions along the dendrites~\cite{2002Migliore}. All of these mechanisms generate non trivial phase shifts to the signals arriving at the somas.

This work enters in this context to analyze the possibility of the CA1 spikes generation resulting from the synchronisation of the signals arriving from a CA3 neuron, through synapses distributed at two different distances form the CA1 soma. Synchronisation is studied for different biological parameters such as the distance of synapses from the soma, the contribution of inhibitory interneurons, the delay in the inhibition, and the activation delay of the proximal synapse distribution. The synapses here considered are AMPA-mediated, having fast rise and decay time, and are present in all neurons in a mature stage of their formation in brain.

Additionally, the analysis considers the background activity on the somas to naively account for large network contributions, as seen in {\it in vivo} neural cells. 

For this purpose, we apply HH dynamics to two identical neurons subjected to identical Poissonian current inputs activating all the synapses at the two group of distances considered. We study synchronisation through a newly defined measure for spiking correlation, using the concept of phase difference between consecutive spikes.

The article is organised as follows: In Sec.~\ref{methods} we provide an explanation of the model, along with the correlation measure defined to assess the synchronisation between spikes. Section \ref{results} contains the main result for the parameters considered, and Section \ref{stoch} analyses the background contribution to the somas. The paper concludes with some final considerations in Sec.~\ref{conclusions}.

\section{Methods} \label{methods}

\begin{figure}[]
 \centering
 \includegraphics[width=0.95\linewidth]{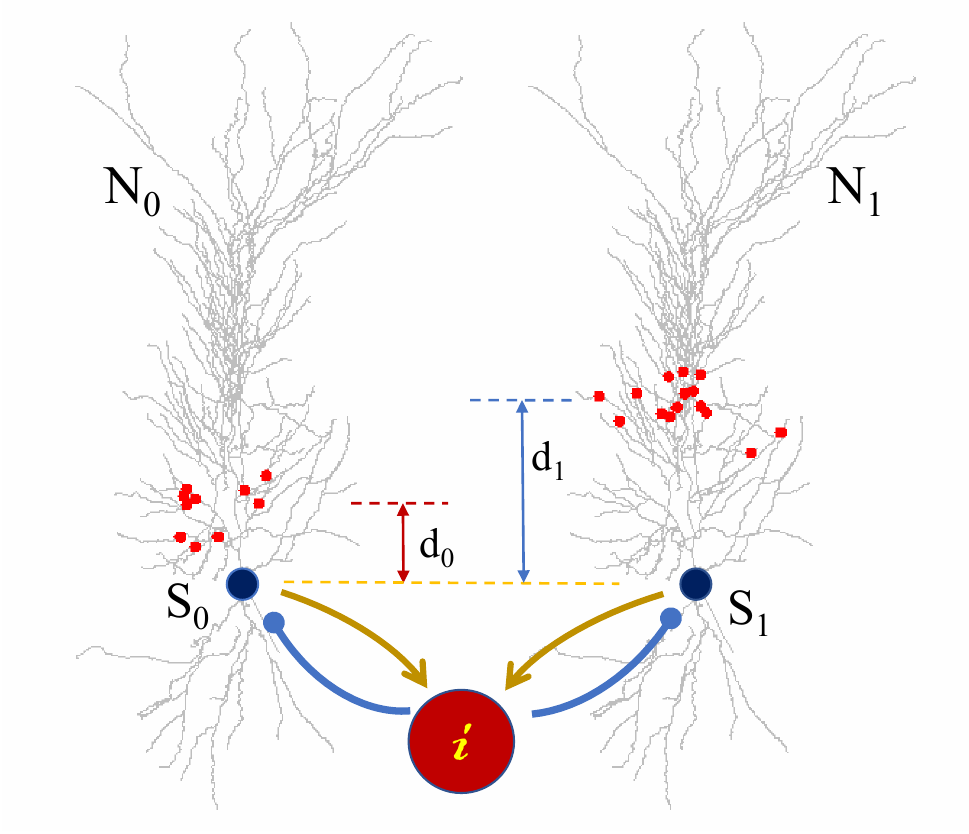}
 \caption{Scheme of the two identical CA1 neurons, specifically the neuron labelled mpg141017\_a1-2\_idC~\cite{2018PLoS_migliore}, that receive inputs from the synapses (red dots in the figure) randomly distributed at different distances from the soma along the apical dendrites. The left neuron $N_0$ receives the closer synapse distribution ($d_0$) to the soma, whilst the right one ($N_1$) receives the synapses in a more distal region ($d_1$). The model includes an interneuron ($i$) which, activated by the somas spiking, acts in inhibiting both neuron activity.}
\label{Neurons}
\end{figure}
The idea is to study the spiking activity on the CA1 neurons, when the signals arrive from two different paths. For this, we use two identical neurons and generate over them the input as provided by synapses located at two different group positions. The spikes are generated as pulses of currents with a Poissonian distribution times. The source ---unique for the two neurons--- is connected to the target by means of $N_{\rm syn}=20$ synapses that are randomly distributed at a certain mean distance from the soma and vary in each numerical experiment. The chosen $N_{\rm syn}$ represents a reasonable number of active synapses at a certain time \cite{2001Megias}.
 
Fig.~\ref{Neurons} shows two identical realistic and optimised neurons (see~\cite{2018PLoS_migliore}) that receive their inputs from the same axon coming from the CA1 neurons of the trisynaptic path. The synapses here considered AMPA-mediated. This means that their rising current responses are fast, as well as their decaying time. Moreover, the position of the synapses is not  arbitrary. The arriving synapses are located in the apical dendrite area, with a random distribution with a mean value close to the soma in neuron $N_0$, and with a greater distance from the soma in neuron $N_1$ (See Ref.~\cite{2008Spruston} for a review on hippocampal neuron morphology). 

\subsection{Equations and model details.}

The microscopic neuron dynamics is based in the Hodgkin-Huxley set of equations, implemented in the NEURON simulation environment \cite{1997HinesCarnevale}, which are freely distributed by Yale university.

The HH equations are based on the electrodynamics of the neuron membrane and gates, characterized by specific excitability that is reflected in their conductances. 

In order to refresh the main mathematics involved in this paper, we sketch rapidly the original equations. The membrane potential of the neuron in the neighbourhood of the channel $j$ has been pioneeringly described by HH equation as follows.
\be
C_{\rm m}\dot{V}_{\rm j} = I_{\rm j}^{\rm ext} - I_{\rm j}^{\rm m} - I_{\rm j}^{\rm syn},
\ee
where $I^{\rm ext}$ is the external current administered to the gate in the membrane area (along this work this current is always zero), $I^{\rm m}$ is the membrane current generated in the membrane, and $I^{\rm syn}$ is the postsynaptic current due to the signal received by the incoming synapses.
The membrane current was schematically described by HH as
\be
  I_{\rm j}^{\rm m} = g_{\rm Na} m^3 h(V_{\rm j} - V_{\rm Na}) + g_{\rm K} n^4 (V_{\rm j}- V_{\rm K} ) + g_{\rm L} (V_{\rm j} - V_{\rm L}),
\label{i_HH}
\ee
where $g_{\rm a}$ (a = Na, K, L) represent the maximum conductance for the ionic currents of sodium, potassium and passive channel (L), respectively, while $V_{\rm a}$s are the equilibrium potentials of the corresponding channels. 
The \emph{gating variables} $m$, $h$, and $n$ represent the ratio of activation and inactivation of the sodium channels and the activation of the potassium channels, respectively. 

The considered ion channels are voltage-gated and their opening follows the differential equation
\be
  \tau_{\rm q}(V) \frac{dq}{dt}= q_{\infty}(V) -q.
\ee
with $q$ representing a generic gating variable; the time function can be expressed as $\tau_{\rm q}(V) = \frac{1}{\alpha_{\rm q}(V)+\beta_{\rm q}(V)}$, and $z_{\infty}(V) = \frac{\alpha_{\rm q}(V)}{\alpha_{\rm q}(V)+\beta_{\rm q}(V)}$ 
where the rates $\alpha_{\rm q}$ and $\beta_{\rm q}$ are fitted from experimental data in a sigmoidal fashion due to their underlying kinetic reaction~\cite{1994Destexhe}.

Beside this simple summary of the equations, the realistic model explored here imposes a more complex equation setting than Eq.~\ref{i_HH}. In fact, this study considers an extended number of channels which takes into account the experimentally known features of CA1 pyramidal cells, as discussed in recent papers on hippocampal neurons~\cite{2010Ascoli,2010Morse}, against which the channels configuration and distribution have been thoroughly validated. The complete set of active membrane properties includes the essential sodium current (Na), four types of potassium (K$_{\rm DR}$, K$_{\rm A}$, K$_{\rm M}$, and K$_{\rm D}$), three types of Calcium (CaN, CaL, CaT), the nonspecific I$_{\rm h}$ current, and two types of Ca-dependent K$^+$ currents, K$_{\rm Ca}$ and C$_{\rm agk}$. 

With all these contributions, the membrane current term of Eq.~\ref{i_HH} modifies in
\be
  I_{\rm j}^{\rm m} = \sum_a g_a f_a(m,h,n)  (V_{\rm j} - V_a),
\label{i_HHa}
\ee
where the sum is extended over all the considered channels, $g_a$ is the maximum conductance, $f_a$ the gating variable, $V_{\rm j}$ the membrane potential and $V_a$ the reversal potential for the channel $a$, where $a$= (Na, K$_{\rm DR}$, K$_{\rm A}$, K$_{\rm M}$, K$_{\rm D}$, CaN, CaL, CaT, I$_{\rm h}$, K$_{\rm Ca}$, C$_{\rm agk}$). More details on the specific shape of the functions $f_{\rm a}$ and the values of  $g_{\rm a}$ can be found in Ref.~\cite{2010Ascoli}.

All dendritic compartments present a uniform distribution of channels, except for K$_{\rm A}$ and I$_{\rm h}$ which in pyramidal cells are known to increase linearly with the distance from the soma~\cite{1999Hoffman,1999Magee}. The values for the peak conductance of each channel were independently optimised in each type of neuron compartment (soma, axon, basal and apical dendrites), with difference of one order of magnitude~\cite{2018PLoS_migliore}.

Concerning the postsynaptic current $I^{\rm syn}$, this is modelled for each synapse as~\cite{1994Destexhe}
\be
  I_{\rm j}^{\rm syn} = -g_{\rm syn} \alpha(t-t_{\rm spk}-\tau_{\rm spk}) (V_{\rm j} - E_0) ,
\label{i_syn}
\ee
with $g_{\rm syn}$ the maximum synaptic conductance. The rate is a double exponential $\alpha(t) = \frac{1}{\tau_{\rm d} - \tau_{\rm r}} (e^{-t/\tau_d} - e^{-t/\tau_r})$, where the mean rise time $\tau_{\rm r}$ and the decay time $\tau_{\rm d}$ are immediately recognised, and the time is resettled after every spike ($t_{\rm spk}$), while $\tau_{\rm spk}$ represents a refractory period in which the neuron remains insensitive. $E_0$ is the synaptic reversal potential, that takes the values $E_0=0\,{\rm mV}$ in excitatory synapses, and $E_0=-80\,{\rm mV}$ in inhibitory ones. The different reversal potentials are responsible for the sign change of the current induced by the two types of synapses in the dendritic membranes, thus allowing the inhibitory synapses ---activated with a suitable physiological delay after the arrival of the stimulus in the circuit--- to act as fast attenuators of the membrane potential generated by the excitatory synapses. In this way, the contribution of the inhibitory synapses plays a crucial role in the possibility that the propagation of the signal can generate a depolarisation in the soma.

Our simulations use $\tau_{\rm r}=3\,{\rm ms}$, and $\tau_{\rm d}=5\,{\rm ms}$, which makes our excitatory synapses of fast AMPA-mediated type~\cite{1999Kleppe}, while the inhibition rate is modeled as a single exponential with decay time $\tau_{\rm I,d}=30\,{\rm ms}$.

The presynaptic Poissonian source is generated with a mean frequency $f_{\rm inp} =60 \,{\rm Hz}$ corresponding to a gamma brain activity.  
The two groups of synapses are located as follows: the proximal ones at a fixed mean distance $d_0=100\,{\rm \mu m}$ from the soma on neuron $N_0$, and the other on $N_1$ at a variable distance $d_1 \in [100,300]\,{\rm \mu m}$. 
The conductance weights of the $N_{\rm syn}$ excitatory synapses are considered to have the same value $sw=5\times 10^{-4}\,{\rm nS}$, whilst the weight of the inhibitory synapses spans in the range $iw \in[0,0.06]\,{\rm nS}$.

To smooth the stochasticity of the outcomes  and reduce the fluctuations of the curves, the results of our simulations have been averaged over $N_{\rm exp}=100$ realisations.

\subsection{Phase synchronisation}
The correlation between spikes in the two neurons is studied by means of a novel measure $c$, which modifies an equivalent measure already presented in~\cite{2011Perez}. The idea is to define the phase of a spiking neuron as~\cite{1997Pikovsky}:
\be
	\phi_{\rm i}(t) = 2\pi\frac{t-t_{\rm k}}{t_{\rm k+1,i}-t_{\rm k,i}},
\ee
where the index $i=0,1$ indicates the neuron $N_i$, and the time $t_{\rm k,i}$ is the time of the $k$-th spike of $N_i$. Thus, at time $t$, the neuron lies at a certain fraction of the total time between two successive spikes, the latter covering a phase spanning between 0 and $2\pi$. This allows us to to define the time average of the cosine of the phase difference between the two neurons in a single  trace $j$ as 
\be
	c_{\rm j} = \frac{1}{t_{\rm fin}-{\rm t_{in}}} \int_{t_{\rm in}}^{t_{\rm fin}} \cos (\phi_1(t) - \phi_0(t)) \,dt,
\ee
where $t_{\rm fin}-t_{\rm in}$ defines the maximum window used for the time average in which the phase results well-defined for both neurons.
This measure is normalised in the interval $[-1,1]$, where the value $-1$ represents the anti-correlated spiking (phase difference equal to $\pi$) and the full correlation is given by $+1$ (phase difference 0). The random distribution of phases is then revealed by the correlation value $c_{\rm j} = 0$ (mean phase difference $\pi/2$).

Given the stochastic nature of the spiking, this measure is then averaged over the number of realisations simulated $N_{\rm exp}$:
\be
	\langle C_{\rm R} \rangle = \frac{1}{N_{\rm exp}} \sum_{j=1}^{N_{\rm exp}} c_{\rm j},
	\label{cmmean}
\ee
This spiking synchronisation measure, or others of the same kind, can be generalised for neuron synchronisation in large networks~\cite{2011Perez}, though for pairwise correlations other options are also used~\cite{2007DelaRocha}.

\section{Results} \label{results}

\begin{figure}[t]
 \centering \includegraphics[width=\linewidth]{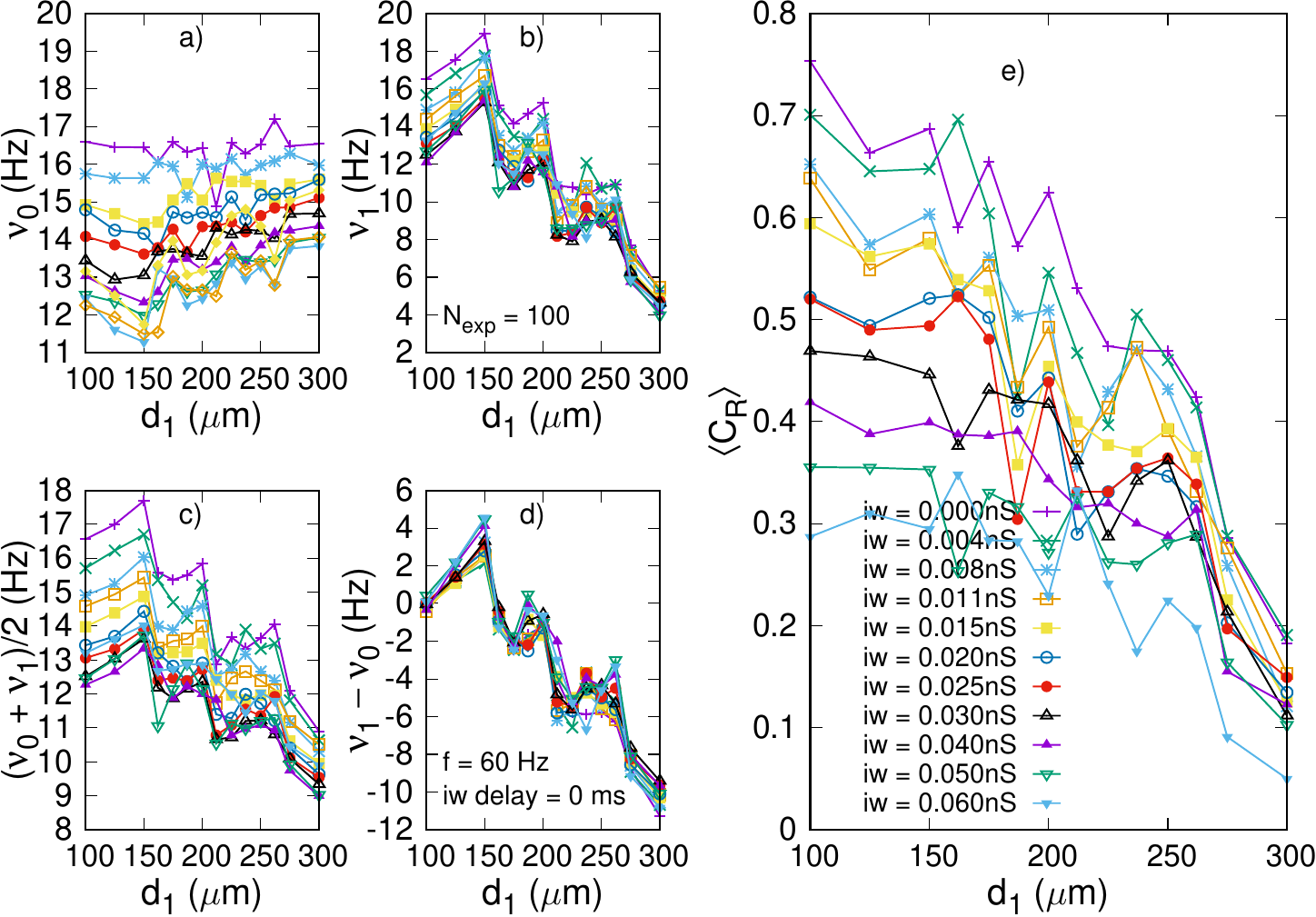}
 \caption{Frequency response of the two neurons $\nu_0$ (panel a) and $\nu_1$ (panel b) to the Poissonian input at the average frequency $f_{\rm inp}=60\,{\rm Hz}$ as a function of the distance $d_1$ of the synapses from the soma in neuron $N_1$. Panel c) shows the mean frequency between the two neurons, and panel d) shows their difference. The spiking correlation (panel e)) shows a monotonic decrease with $d_1$.}
\label{Synch_d1}
\end{figure}

Figure~\ref{Synch_d1} shows the frequency response of the two neurons $\nu_0$ and $\nu_1$ (panels a) and b)) under a frequency input $f_{\rm inp}=60\,{\rm Hz}$ as a function of the distance $d_1$ for different inhibitory weights $iw$. 
We observe a weak variation of $\nu_0$ as a function of $d_1$, with a nonmonotonic trend. Surprisingly, although for increasing  values of the $iw$ parameter the spiking activity decreases, we note an increase for increasing distances $d_1$ with a saturating trend. This means that the farther is the second groups of synapses from the soma, the lower {tends to be} their effect on $N_0$. These behaviours show how rich the spiking phenomenon can be when considering realistic neuron morphologies, and how relevant the role of the interneuron is in mediating the interaction between pyramidal neurons.

An interesting result consists of the slight increase in spiking activity for the neuron $N_1$ as the distance $d_1$ increases. In fact, panel $b)$ shows an increase in $\nu_1$ up to a distance of $d_1^*=150\,{\rm \mu m}$. For greater distances, the spiking activity of $N_1$ tends to decrease with a non-monotonic behaviour, probably due to the highly structured dendritic distribution of the realistic neuron. We also note that around the same distance value $d_1^*$, the spiking activity of the two neurons inverts in intensity, as visible in the negative value of the $\nu_1-\nu_0$ magnitude (panel d), for $d_1 > d_1^*$.
The sawtooth shape visible in the spiking frequency in Fig.~\ref{Synch_d1}(b-d) is only due to the specific morphology used. In fact, some calculations made by using a different neuron (not shown), do not present the complex curve structure here found.

The phase spiking correlation $\langle C_{\rm R} \rangle$ presented in panel e) presents a monotonic decrease as a function of both $d_1$ and the inhibition weight $iw$ (See Fig.~3 below). This result seems quite natural. This result seems quite natural. In fact, it is expectable that the signals generated in more distal synapses (in $N_1$) will reach the soma reduced in amplitude and smoothed over a wider time envelope and, consequently, they result in a decrease in the spiking correlation. 

\begin{figure}[b] \centering
 \includegraphics[width=\linewidth]{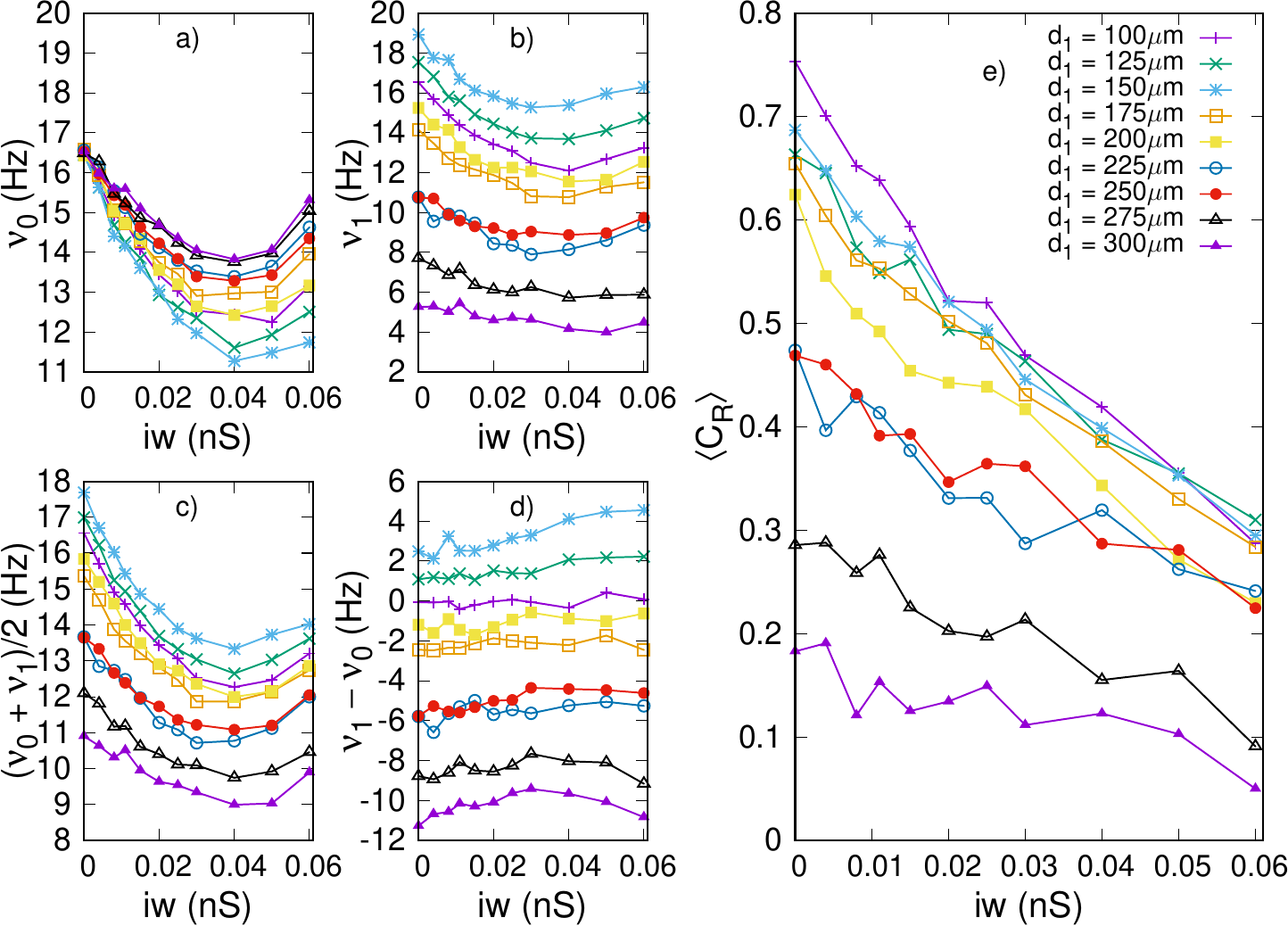}
 \caption{Frequency response equivalent to the one shown in Fig.~\ref{Synch_d1}, shown as a function of the inhibitory weight $iw$. The delay in the inhibition response is here 0ms.}
\label{Synch_iw}
\end{figure}

To better visualise these results, Fig.~\ref{Synch_iw} presents the same calculations as Fig.~\ref{Synch_d1}, replotted as a function of the inhibition weight $iw$. The frequency response of both neurons (panels a) and b)) as well as their average (panel c)) show clear non-monotonic behaviour of spiking frequency, with a minimum of $\nu_0$ around $iw^* \approx 0.04\,{\rm nS}$, which corresponds to about 40 active inhibitory synapses.

Surprisingly, the synchronisation parameter $\langle C_{\rm R} \rangle $ appears monotonically decreasing with $ iw $, and does not reflect the non-monotonic behaviour evident in the mean spiking frequencies.
This result appears in contradiction to the well-known fact that a small amount of inhibition improves the spiking synchronisation between neurons ~\cite{Jasper,2010Bazhenov}, and the topic has been addressed in a paper (Fiasconaro and Migliore, under revision), where a series of calculations with realistic neurons have shown that, under a stochastic current source, an increase in synchronisation with inhibition is obtained, provided that the inactivation times of the excitatory synapses $\tau_{\rm d}$ are long enough. This is not the case in this study, since $\tau_{\rm d} < \tau_{\rm I,d}$.

\begin{figure}[tp]
 \centering
 \includegraphics[width=\linewidth]{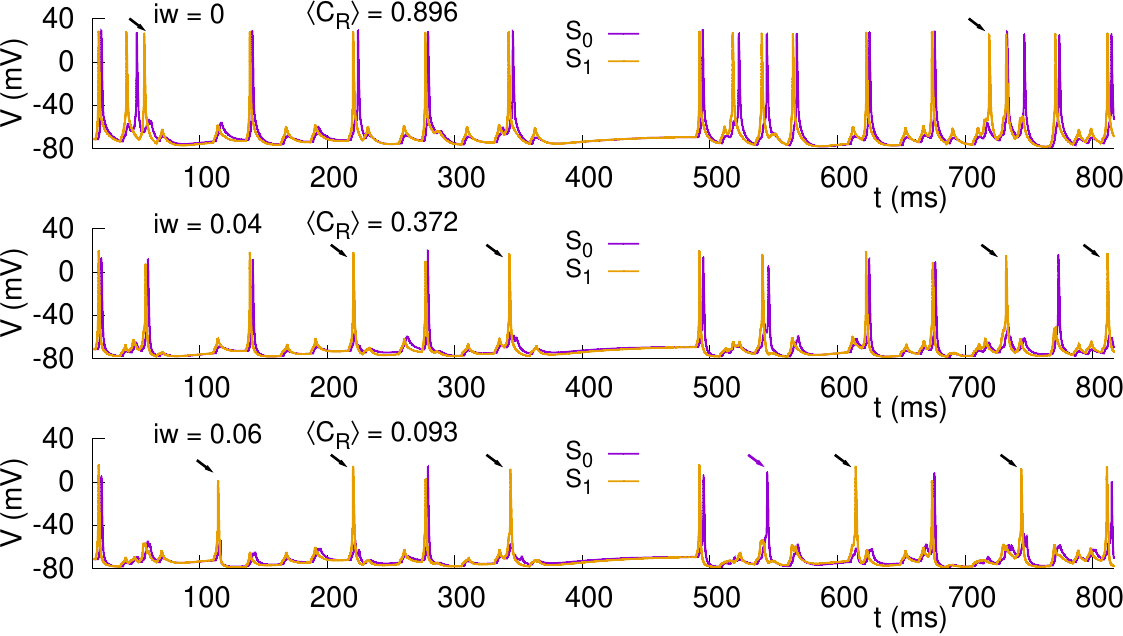}
 \caption{Time evolution of the spiking activity in the two neurons with similar (on average) synapse locations ($d_0=100$ and $d_1=100\,{\rm \mu m}$) at $f_{\rm inp}=60\,{\rm Hz}$ for $iw=0, 0.04, 0.06\,{\rm nS}$. It is evident that there is a frequent anticipation of soma spikes in $S_1$ compared to $S_0$, and a decrease in frequency response as $iw$ increases, also due to spiking failures. The trace $S_0$ has been shifted forward by 2\,{\rm ms} to avoid the complete superposition in some cases. The tiny arrows indicate the spikes occurring only in one of the two somas.}
\label{Synch_traj_ant}
\end{figure}

Fig.~\ref{Synch_traj_ant} presents the spiking traces which correspond to the three cases: $iw=0, 0.04, \,{\rm and}\, 0.06\,{\rm nS}$. There, another counterintuitive and interesting occurrence is visible, i.e. the anticipation of the spiking of the neuron occurring with distal synapses with respect to those with the proximal ones. In fact, the spikes shown in the time window are always characterised by the anticipation of the spikes in $S_1$, with synapse distribution at $d_1=100\,{\rm \mu m}$, in comparison to the spikes of $S_0$, with more proximal synapses ($d_0=100\,{\rm \mu m}$). In the plot, there is a forward time shift of 2\,ms for visualisation purposes, so the effect is just slightly amplified, but not hidden in the figure. This behaviour depends on the fact that distal dendrites are narrower than proximal ones, and it has been demonstrated that the depolarising propagation performs better and more efficiently there than in wider dendrites~\cite{2005MiglioreAscoli}. In fact, the rebound of the current in the dendrite edges, for a suitably short distal extreme of the branch, can allow an overlap and integration of ion currents inside the dendrite that can more intensely and efficiently stimulate a spiking arrival to the soma, even anticipating the depolarization. This phenomenon can also increase the entire spiking frequency, thus explaining the increase of the frequency response $\nu_1$ for $d_1 < d_1^*$ shown in Fig.~\ref{Synch_d1}b).

Fig.~\ref{Synch_traj_ant} also evidences the non-monotonic response frequency when increasing the inhibitory weight $iw$, and the greater difference in $S_0$ (blue line) compared to $S_1$. Additionally, the spiking response when increasing the inhibition weight is accompanied by more frequent spiking failures in $S_0$ than in $S_1$.

In order to better understand the role of some time delays in CA1 synchronisation, we performed a set of calculations by changing the synaptic inhibition delay $\tau_{\rm I,s}$, i.e. the delay employed by the interneurons to react and inhibit both neurons.
\begin{figure}[bp]
 \centering
 \includegraphics[width=\linewidth]{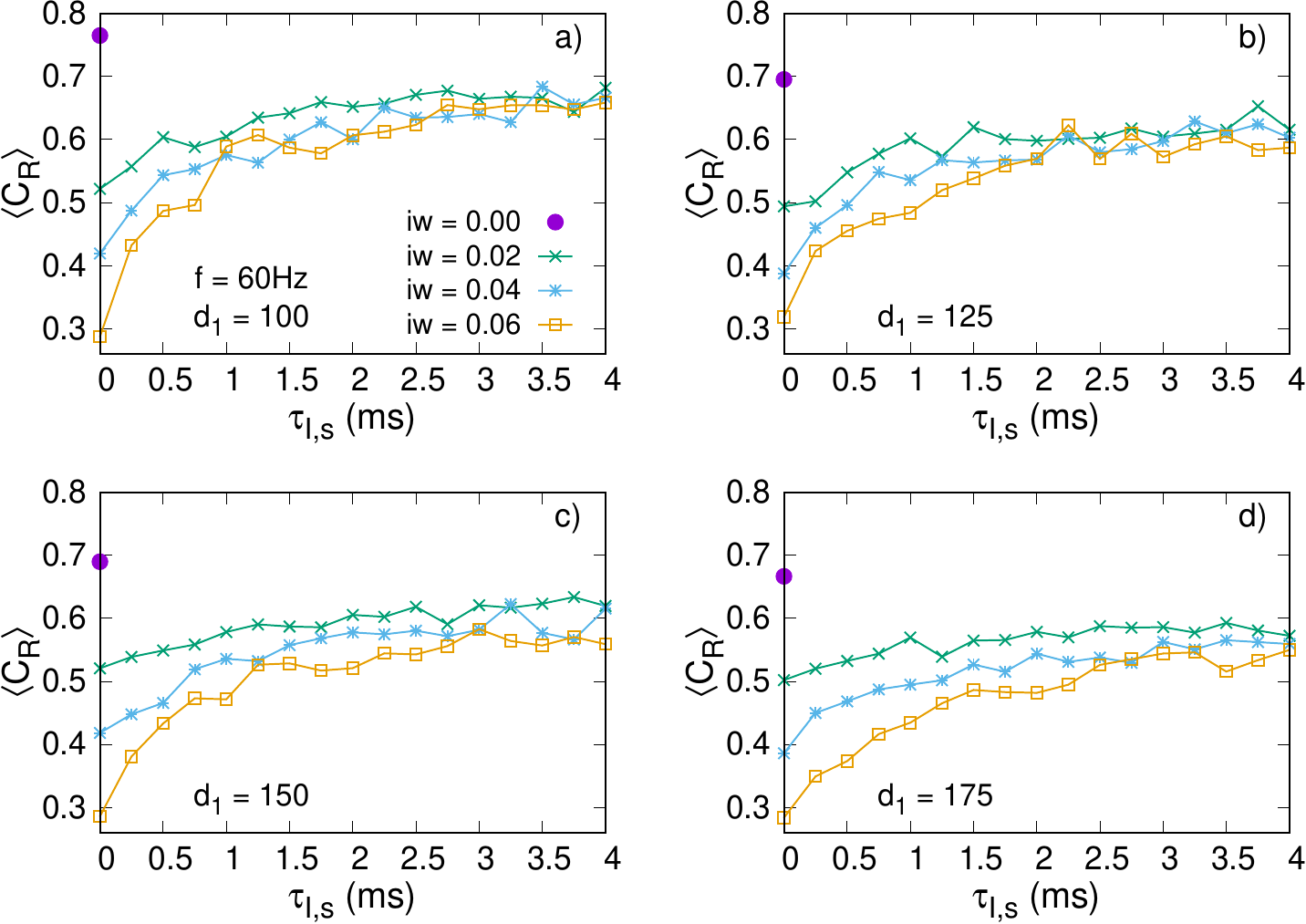}
 \caption{Spiking correlation is shown as a function of the synapse inhibitory delay $\tau_{\rm I,s}$ for various values of the inhibitory weight, specifically $iw=0, 0.02, 0.04, 0.06\,{{\rm nS}}$. The panels refer to different mean distances from the soma of $N_1$, $d_1 =100, 125, 150, 175\,{\rm \mu m}$. The presence of inhibition ($iw\neq0$) reduces the spiking synchronisation $\langle C_{\rm R} \rangle$ in comparison to the uninhibited values (represented by a full circle in the plots). For inhibited dynamics, $\langle C_{\rm R} \rangle$ increases monotonically with a saturating trend, almost reaching the uninhibited value.}
\label{Synch_delay}
\end{figure}
Fig.~\ref{Synch_delay} shows the phase spiking correlation $\langle C_{\rm R} \rangle$ as a function of the synaptic delay $\tau_{\rm I,s}$ in the inhibition of the interneurons' activation by the soma's spiking.  The inhibition acts with both a forward and backward mechanism, because, regardless which neuron spikes, the interneurons act on both neurons at the same time (see the graph in Fig.~\ref{Neurons}). 

The interneuron has not been explicitly implemented in the network with a realistic morphology, but it is modelled as an inhibiting synapse's current according to Eq.~\ref{i_syn}. For this reason, to simulate a realistic signal propagation in it, we introduced a delay in which the inhibition starts to understand its role in the synchronisation.

The plots indicate that the presence of inhibition decreases the synchronisation features. In fact, all the curves (with $iw=0.01, 0.02, 0.05\,{\rm nS}$) lie below the full stop in the figures, which represents the absence of inhibition ($iw=0\,{\rm nS}$). Specifically, the absence of delay represents the maximum loss of synchronization, while a saturating increase of synchronization is reported for increasing values of $\tau_{\rm I,s}$. The same behaviour is confirmed for all the distances from the soma in the neuron $N_1$ shown in the three panels of Fig.~\ref{Synch_delay}, specifically $d_1=100\,{\rm \mu m}$, i.e. no mean distance between the synaptic location of the two neurons, $d_1=150\,{\rm \mu m}$ and $d_1=200\,{\rm \mu m}$. The difference in the three distances only reflects the expectable outcome that by increasing the distance $d_1$, the synchronisation decreases, as shown in Fig.~\ref{Synch_d1}.

In order to study how some differences in the path followed by the signals arriving at the CA1 from the CA3 neurons can affect synchronisation, a series of simulations have been performed in which the excitation input on the closest synapse group of the CA1 neuron has been delayed a few milliseconds with respect to the distal group, thus simulating a longer path previously followed by the axons arriving to this group of synapses. This \emph{path delay} ($\tau_{\rm E,del}$) has been applied to the activation synapses of neuron $N_0$, with the aim of simulating the possible correlation improvement that compensates current delays of the distal synapses.

\begin{figure}[tb]
 \centering
 \includegraphics[width=\linewidth]{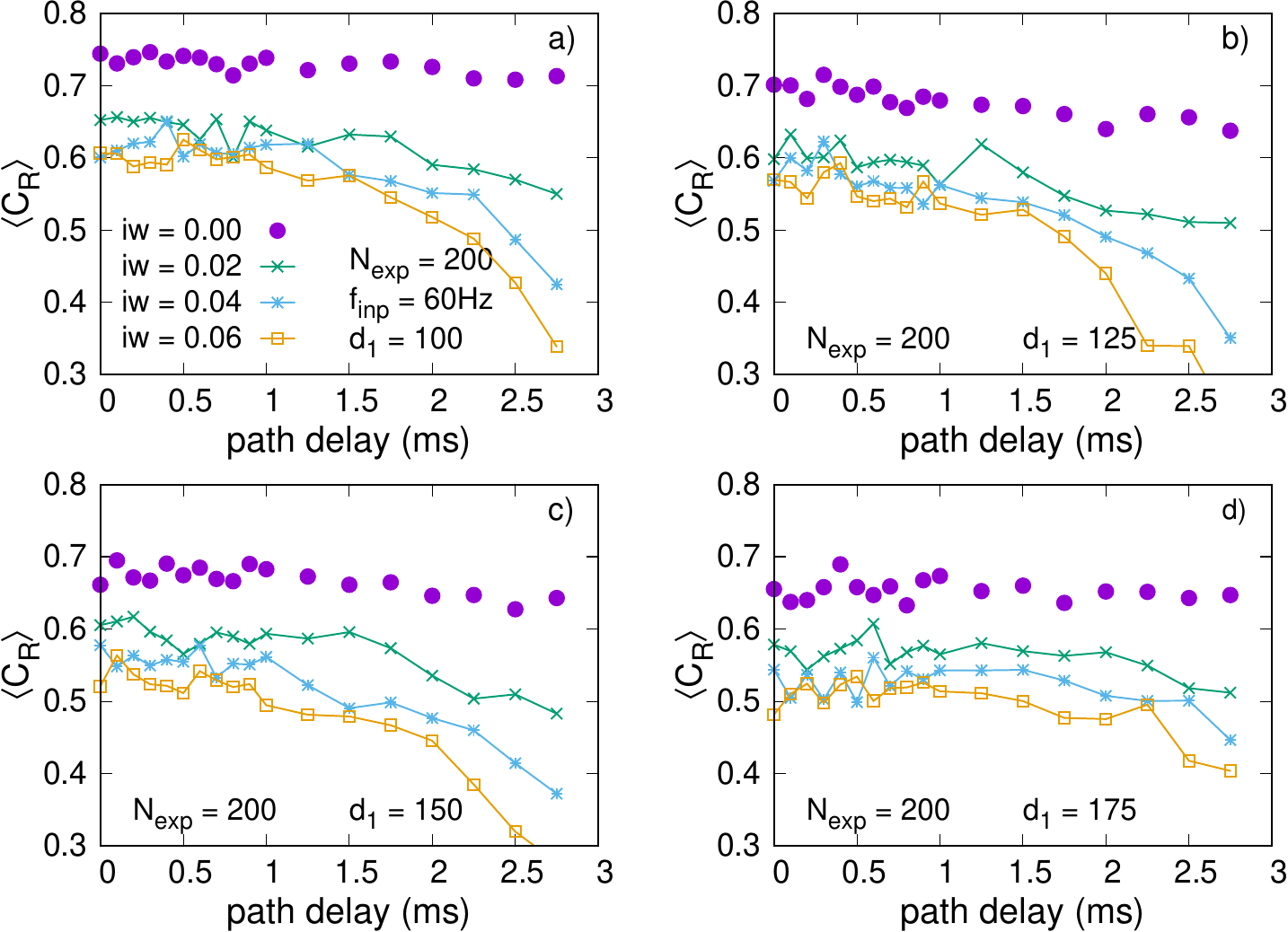}
 \caption{Spiking correlation as a function of the \emph{path delay}  $\tau_{\rm E,del}$ for different inhibition weights and different distances $d_1$ from the soma. Synchronisation appears weakly affected up to values of  $\tau_{\rm E,del} \approx 2\,{\rm ms}$. The inhibition delay here is $\tau_{\rm I,s}=2\,{\rm ms}$.}
\label{Synchro_DD_M1-2D}
\end{figure}

Fig.~\ref{Synchro_DD_M1-2D} shows the spiking correlation $\langle C_{\rm R} \rangle$ as a function of this delay for four inhibition weights ($iw$ =0, 0.01, 0.02 and $0.05\,{\rm nS}$) and four distances of the second neuron from the soma: $d_1=100,125,150,175\,{\rm \mu m}$.
We can observe here that no significant contribution is due to the excitatory delay $\tau_{\rm E,del}$ in synchronising the signals for the different inhibition weight explored. In fact, though for all the distances the curves with $iw\neq0$ lie below the full blue dots indicating the correlation with no inhibition, those curves seem to remain essentially unmodified for low time delays ($\tau_{\rm E,del} \lesssim 1.5\,{\rm ms}$), time at which a fast synchronisation decay occurs. This calculation confirms that the presented model predicts that the regulatory function of the interneurons with their inhibitory activity on the CA1 neurons barely affects AMPA-mediated synapses.

\section{Stochastic Background contribution.} \label{stoch}

The above described model has been analysed here as the two neurons were disconnected from the brain network, only connected to the incoming signals modelled as a Poissonian process activating the excitatory synapses. In order to better understand a more complex and realistic scenario, we performed a series of calculations where, maintaining the distributed synapses already investigated in the previous section, the two somas are reached by a noisy background, with the aim of representing in a simple way the stochastic contribution provided by the many connections of the brain networks on the activity of the two neurons studied here, thus reproducing a {\em in vivo}-like activity.
This contribution follows the Destexhe approach~\cite{1999Destexhe,2001Destexhe} where two synapses, one excitatory and the other inhibitory, are both characterized by a mean value of the maximum conductances $g_{\rm OU,x}$ along with an additional stochastic term. This latter is modelled as an Ornstein-Uhlenbeck (OU) fluctuation with a certain noise intensity $\sigma_{\rm x}$ and a correlation time $\tau_{\rm OU,x}$, where the suffix $x$ indicates the two possibilities: $x = E$ or $x = I$ for the excitatory synapse or the inhibitory one, respectively. With this ``fluctuating conductance approach" they were able to successfully simulate, under a simpler than that presented in this paper, a neocortical neuron response that reproduced a typical experimental trace.

The equation added to the system is then the two-term contribution to the current 
\be
  I^{\rm stoch} = g_{\rm OU,E} (t) (V - E_{\rm E}) + g_{\rm OU,I} (t) (V - E_{\rm I}),
\label{i_stoch}
\ee
with reversal potential $E_{\rm E}=0$ and $E_{\rm I}=-80$, values similar to the distributed synapses. The maximum conductances follow independently of each other an Ornstein-Uhlenbeck process:
\be
  \frac{dg_{\rm OU,x} (t)}{dt}  = \frac{1}{\tau_{\rm OU,x}} [ g_{\rm OU,x} - g_{\rm x} ] + \sqrt{2\sigma_{x}} \eta_{\rm x}(t),
\label{g_stoch}
\ee
where $g_{\rm x}$ are the average maximum conductances of the added synapses, $\tau_{\rm OU,x}$ is the correlation time of the OU process, $\sigma_{\rm x}$ is the intensity of the fluctuations, and $\eta_{\rm x}(t)$ is the Gaussian noise that is uncorrelated in time with zero mean ($\langle \eta_{\rm x}(t) \rangle =0$ and correlation $\langle \eta_{\rm x}(t)\eta_{\rm y}(t') \rangle =\delta(t-t')$). Please refer to \cite{1996Gillespie, 2001Destexhe} for further details on the updating rules when numerically integrating the above equations.

We performed three different sets of calculations. In the first, we only spanned some conductance values and corresponding noise intensities with the excitatory term, excluding the inhibition contribution. In the second, we only spanned the inhibitory term, again in mean conductance and intensity of fluctuations, by excluding the excitatory contributions. In the third one, we spanned the combined contributions of inhibition and excitation, fixing the value of the two corresponding intensities of fluctuation to the value of the maxima registered in the previous cases. The OU time correlations have been fixed to the values $\tau_{\rm OU,E}=3\,{\rm ms}$ in the case of excitatory synapses, and to the value $\tau_{\rm OU,I}=10\,{\rm ms}$ in the case of inhibitory ones.

\begin{figure}[htbp]
 \centering
 \includegraphics[width=\linewidth]{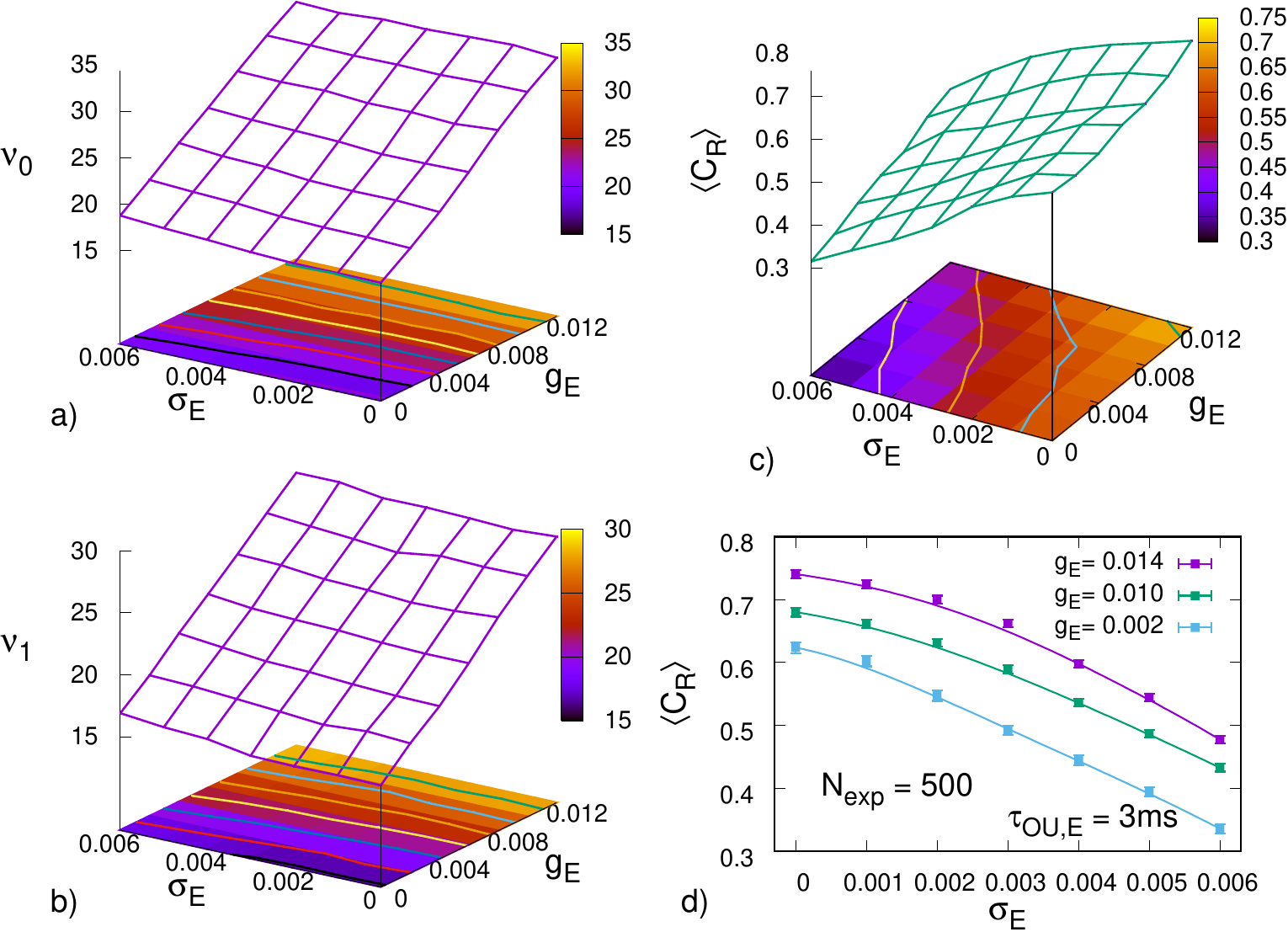}
 \caption{Spiking correlation as a function of the excitatory fluctuating conductance parameters: the average conductance $g_{\rm E}$ and noise intensity $\sigma_{\rm E}$. The OU correlation time is $\tau_{\rm OU, E} = 3\,{\rm ms}$ in this case. The parameters for the distributed synapses in this plot are $iw = 0.01\,{\rm nS}$, $d_1= 175\,{\rm \mu m}$, and $\tau_{\rm d}=5\,{\rm ms}$. The excitatory fluctuating term decreases monotonically the spiking correlation $\langle C_{\rm R} \rangle$.}
\label{Synchro_Stoch_E_3d}
\end{figure}

\begin{figure}[tb]
 \centering
 \includegraphics[width=\linewidth]{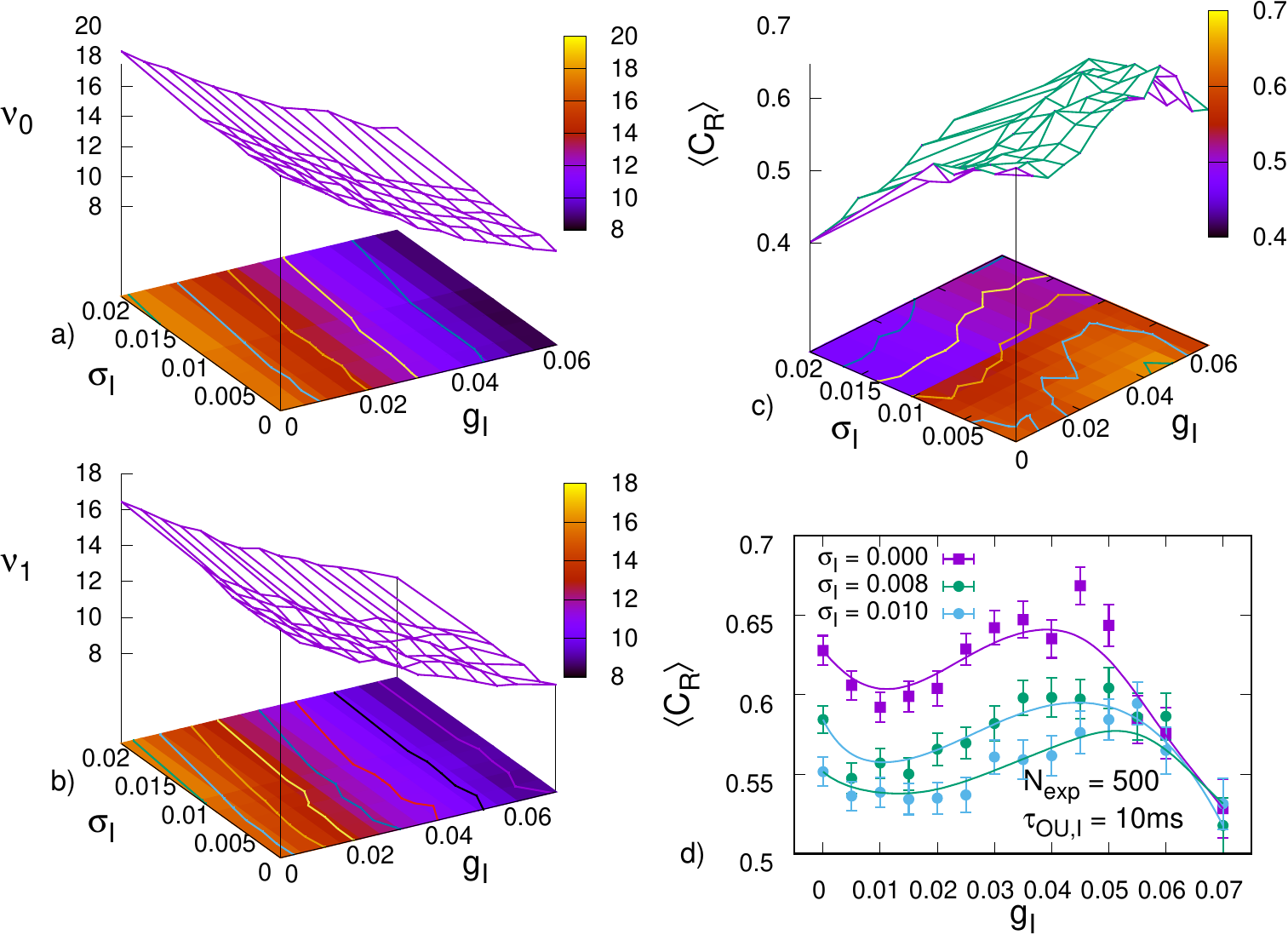}
 \caption{Spiking correlation as a function of the inhibitory fluctuating conductance parameters: average $g_{\rm OU, I}$ and noise intensity $\sigma_{\rm I}$. The OU correlation time is here $\tau_{\rm OU, I} = 10\,{\rm ms}$. The parameters of the distributed synapses are the same as Fig.~\ref{Synchro_Stoch_E_3d}.}
\label{Synchro_Stoch_I_3d}
\end{figure}

\begin{figure}[htb]
 \centering
 \includegraphics[width=\linewidth]{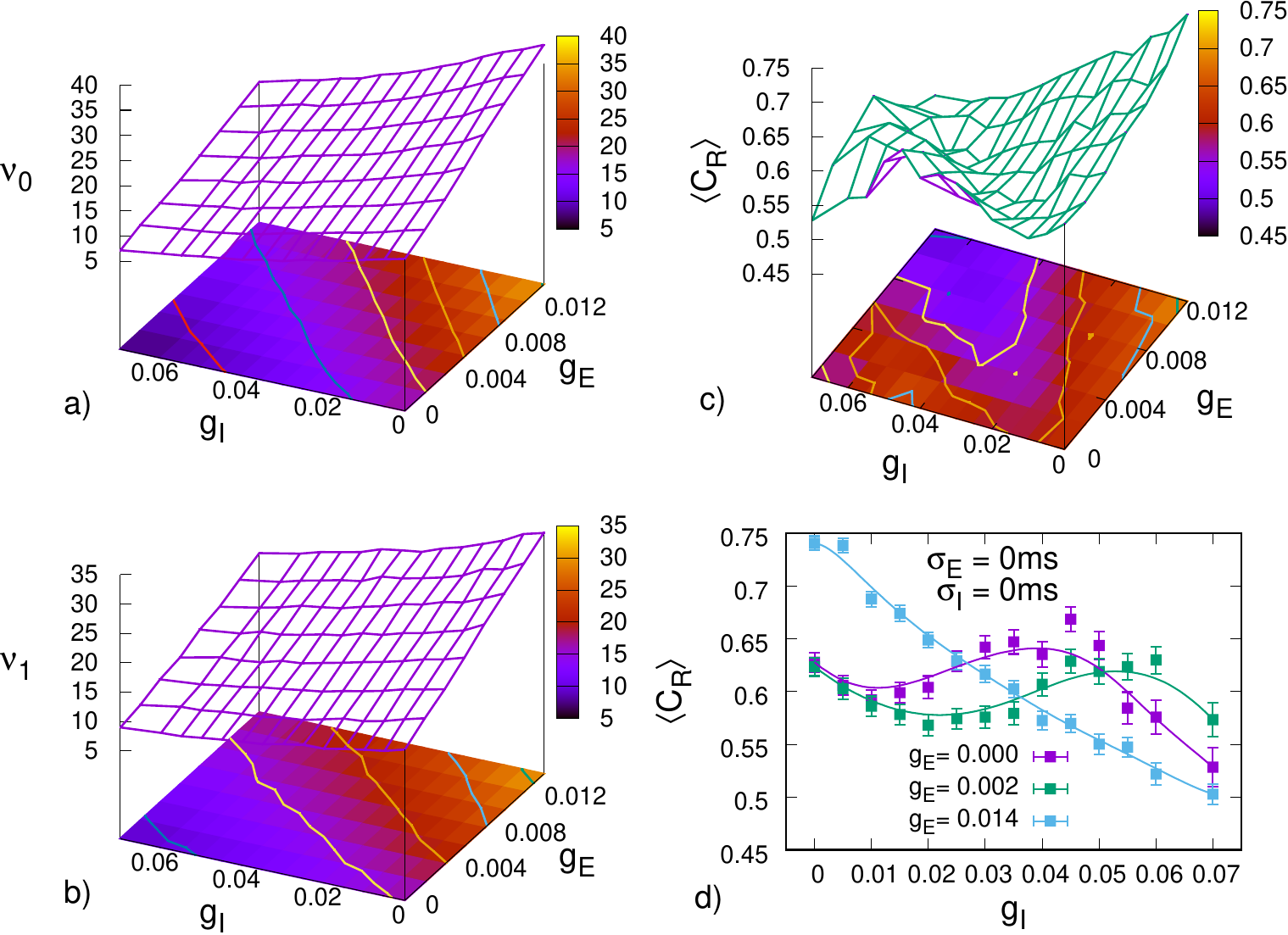}
 \caption{Spiking correlation as a function of both excitatory and inhibitory background conductances in the OU scheme: The parameters of the distributed synapses are the same as Figs.~\ref{Synchro_Stoch_E_3d} and~\ref{Synchro_Stoch_I_3d}.}
\label{Synchro_Stoch_EI_3d}
\end{figure}

Figures~\ref{Synchro_Stoch_E_3d},~\ref{Synchro_Stoch_I_3d} and ~\ref{Synchro_Stoch_EI_3d} show the results of the simulations by spanning the two parameters each time considered and schematized in the panels a) and b), which record the output frequencies of the two somas, panel c) that shows the spiking correlation $\langle C_{\rm R}\rangle$ and panel d) shows the projection of $\langle C_{\rm R}\rangle$ along one of the parameters in the 3D plots.

In Fig.~\ref{Synchro_Stoch_E_3d}a) and b) we can observe how the excitatory-only background activity always increases the average spiking frequency, regardless of the fluctuation intensity. A more structured behaviour is present in panel c) for the spiking correlation, where $\langle C_{\rm R}\rangle$ increases with the mean conductance, whilst a clear decrease is evident as a function of the fluctuation intensities $\sigma_{\rm E}$, the latter results more visible in panel d) for some values of the average maxim conductance $g_{\rm E}$.

More interesting is the response of the neurons to the inhibitory background activity. We can see in Fig.~\ref{Synchro_Stoch_I_3d} that, besides the expected monotonic decrease of the output spiking frequencies in both neurons with the average inhibition weight $g_{\rm I}$, the spiking correlation reveals a non-monotonic behaviour with $g_{\rm I}$ (see panel c)) with a clear maximum that also depends on the noise intensity. Panel d) shows the correlation $\langle C_{\rm R} \rangle$ as a function of the average inhibitory weight $g_{\rm I}$ for three fluctuation intensities, in which a clear maximum is evident with a relative increase with respect to the minimum before of around $\approx 10\%$. Instead, the correlation decreases monotonically with the noise intensity.
Finally, the third set of calculations has been performed with fixed noise intensities in both cases (excitatory and inhibitory) and specifically we have chosen the maximum values of the calculations performed above, i.e. $\sigma_{\rm E}= 0$ and $\sigma_{\rm I}= 0$, but similar behaviour is recovered with non-zero noise intensities. This mixture of excitation and inhibition reveals a clear non-monotonic behaviour of the spiking correlation as a function of both magnitudes $g_{\rm E}$ and $g_{\rm I}$, with optimal correlations for a proper $g_{\rm I}$-$g_{\rm E}$ combination, as seen in Fig.~\ref{Synchro_Stoch_EI_3d}c). Panel d) shows a selection of $\langle C_{\rm R} \rangle$ curves as a function of the $g_{\rm I}$ for three values of the background excitatory weight $g_{\rm E}$, where this complex behaviour of the two measures appears. The balance between excitation and inhibition is an important neurophysiological factor. In fact, the absence of inhibition leads to epileptic activity~\cite{1987Dichter} and loss of sensory selectivity~\cite{1975Sillito}. Some estimates of the weight of the two conductances have been made using data from cortical neurons~\cite{1996Borg-Graham,1998Borg-Graham}, and fitting the output with a LIF model~\cite{2008Monier}, a 4 to 1 ratio between inhibitory/excitatory conductances was calculated. In another case, during slow-wave sleep activity and in depolarised conditions (up-phase), excitatory and inhibitory synaptic conductances are balanced with a ratio of 1\cite{2003Shu,2006Haider}, whereas in the awake state inhibition becomes predominant~\cite{2013Haider}. In the presented model, the inhibitory background activity provides optimal synchronisation conditions at values about one order of magnitude higher than the excitatory ones.

\section{Summary and Conclusions.} \label{conclusions}

We studied the synchronisation behaviour of two CA1 neurons subjected to Poissonian inputs at a frequency $f_{\rm inp}=60 \,{\rm Hz}$ as a function of the different distances from the soma of the excitatory synapses arriving from the CA3 neurons of the trisynaptic path of the Hippocampus.

The spiking activity in the two neurons appear nonmonotonic with the distance of synapse distribution in the second neuron, with a extremal value at $d_1^*$ that is a minimum value for the neuron with the synapse distribution at $d_0$ (See Fig.~\ref{Synch_d1}a)), and a maximum value for N$_1$ with distal synapses (Fig.~\ref{Synch_d1}c)), showing the counterintuitive effect of increase of spiking frequency for distal synapses.

The spiking activity of both neurons decreases nonmonotonically with the inhibitory weight, with a minimum values corresponding to about 40 active interneurons (Fig.~\ref{Synch_iw}).

As concerns the synchronisation features measured by the phase spiking correlation $\langle C_{\rm R} \rangle$ we always register a decrease with the weight of the inhibitory synapses, as well as with the distance of the synapse distribution (Figs.~\ref{Synch_d1} and ~\ref{Synch_iw}). Moreover, in the limit of small distances ($d_1 \lesssim 200\,{\rm \mu m}$) the synchronisation is weakly affected by the path time delays in the excitatory synapses $\tau_{\rm E,del}$ if they remain at low values ($\tau_{\rm E,del} \lesssim 2\,{\rm ms}$) (See Fig.~\ref{Synchro_DD_M1-2D}). 
The absence of increase of synchronization with the inhibition  is essentially due to the short inhibitory time decay of the AMPA-mediated synapses for which the inhibition plays no constructive role. Nevertheless, we also observe, always in the presence of inhibition, an increase of the phase spiking correlation with the time delay $\tau_{\rm I,s}$ at which the interneurons activate (Fig.~\ref{Synch_delay}).

The results presented here show that the fast AMPA-mediated excitatory synapses are not able to increase the synchronisation of soma spiking in the absence of a stochastic background, even in the presence of the inhibition given by interneurons. These latter simply act as modulators, with nonmonotonic behavior, of the frequency response of the neurons. 

Moreover, the presence of the unavoidable stochastic background in \emph{in vivo} functioning (See Figs.~\ref{Synchro_Stoch_E_3d} to \ref{Synchro_Stoch_EI_3d}), reveals that the fluctuations do not play any constructive role in this model, and they mainly contribute to reducing the spiking correlation of the system. Instead, an increase in synchronisation can be achieved by a combined excitatory/inhibitory background activity in the presence of optimal conditions for a range of suitable conductances, with a dominant inhibition weight of about one order of magnitude with respect to excitatory synapses.

\vspace{0.5cm}
\emph{Acknowledgments.---}
The authors acknowledge the grant PID2020-113582GB-I00 funded by MCIN/AEI/10.13039/501100011033, the support of the Aragon Government to the Recognized group `E36\_20R F\'isica Estad\'istica y no-lineal (FENOL)' and the grant from the Swiss National Supercomputing Centre (CSCS) under project ID ich011 and ich002 (to MM), and from the CINECA Consortium (Italy).
We also acknowledge a contribution from the Italian National Recovery and Resilience Plan (NRRP), M4C2, funded by the European Union – NextGenerationEU (Project IR0000011, CUP B51E22000150006, “EBRAINS-Italy”), and the funds of the European Union NextGenerationEU/PRTR through the spanish Ministerio de Universidades (BOA 139 (31185) 01/07/2021).

\end{document}